\documentclass[journal=jpclcd,manuscript=letter]{achemso}
\usepackage[version=3]{mhchem} % Formula subscripts using \ce{}
\usepackage{bm}
\usepackage{graphicx}
\usepackage{amsmath,amssymb}

\title{Photo-induced Molecule Formation of Spatially Separated Atoms on Helium Nanodroplets}

\author{Florian Lackner} 
\email{florian.lackner@tugraz.at}
\author{Wolfgang E. Ernst} 
\affiliation{Institute of Experimental Physics, Graz University of Technology, Petersgasse 16, A-8010 Graz, Austria, EU}

\begin{document}

\begin{abstract}

Besides the use as cold matrix for spectroscopic studies, superfluid helium droplets have served as cold environment for the synthesis of molecules and clusters. Since vibrational frequencies of molecules in helium droplets exhibit almost no shift compared to the free molecule values, one could assume the solvated particles move frictionless and undergo a reaction as soon as their paths cross. There have been a few unexplained observations that seemed to indicate cases of two species on one droplet not forming bonds but remaining isolated. In this work, we performed a systematic study of helium droplets doped with one rubidium and one strontium atom showing that besides a reaction to RbSr, there is a probability of finding separated Rb and Sr atoms on one droplet that only react after electronic excitation. Our results indicate that ground state Sr atoms can reside at the surface as well as inside the droplet. 
 
\end{abstract}

Over the past two decades, superfluid helium droplets have been used as nanocryostats for the study of highly elusive and ephemeral species\cite{Toennies2004,Callegari2011} whose spectroscopy became only possible this way: high-spin alkali trimers,\cite{Higgins1996,higgins1996spin} fragile chains of polar HCN molecules\cite{nauta1999nonequilibrium} as well as the cyclic water hexamers.\cite{nauta2000formation} \\
In our current understanding\cite{Toennies2004,Callegari2011}  of the process underlying the formation of such species it is assumed that, upon doping, the dopants can move freely and frictionless in the superfluid helium environment. Eventually, the dopant atoms or molecules will meet and form a molecular bond. With the generally correct assumption of helium being an only weakly interacting atom, it is often neglected that at the droplet temperature of 0.4~K any small barrier due to the presence of helium may prevent different dopants to find each other.  \\
In this work, we systematically studied a particular reaction that seems to be straight forward but according to our measurements can pursue unexpected pathways. Employing Rb and Sr atoms, we show that formation of RbSr \cite{lackner2014helium,pototschnig2014ab} is not the only outcome. As another result, a substantial number of atoms stay separated from each other and the formation of a molecular bond can be photo-induced by excitation of the Sr atom.  \\
Even though partial coagulation was already considered back in 1995 by Lewerenz $et~al.$\cite{lewerenz1995successive} in order to explain the observed mismatch between the coagulation cross section and the pickup cross section, it has never been considered to be an important process that can have an impact on results drawn from helium droplet isolation spectroscopy. However, some evidences that indicate such a process can be found in literature: The investigation of high-spin alkali dimers and trimers, the formation of which is favored at the surface of helium droplets\cite{Higgins1996,higgins1996spin}, set the stage for helium droplet isolation spectroscopy and sparked a whole new field dedicated to the investigation of these elusive molecules.\cite{reho2001photoinduced,mudrich2004formation,nagl2008heteronuclear,giese2011homo,bruhl2001triplet} Successive capture and coagulation nicely explained the observed spectra, except for a strange observation that raised questions about the underlying formation process: In ref. \citenum{Higgins1998} it has been noted that upon doping of the droplets with Na and K, it is possible, employing laser induced fluorescence spectroscopy, to record a Na$_2$ transition at a K atom emission line. It was speculated that the presence of separated species can explain this observation. Similar thoughts have been expressed later on in ref \citenum{bruhl2001triplet}.  \\
In this context, Mg doped helium droplets (Mg-He$_\text{N}$) represent a special case: For these dopants the presence of spatially separated atoms in form of a foam-like structure has been claimed.\cite{przystawik2008light,hernando2008density} This was explained by the modulation of the long-range Van der Waals part of the Mg dimer potential energy curve by the surrounding helium, causing a potential barrier that hinders the coagulation process. However, Mg has always been considered to be an unusual dopant that behaves differently than others. Later on, speculations about the foam-like structure of Al atoms in helium droplets have been raised\cite{krasnokutski2011low,krasnokutski2015resonant}, which, however, has been highly debated.\cite{spence2014formation} \\
Infrared spectroscopy experiments on Ca and Sr doped droplets co-doped with HCN molecules\cite{douberly2010hcn} revealed other interesting findings: The Ca-HCN molecule that is formed upon pickup has been found at two different sites, on the surface as well as solvated. In contrast, Sr-HCN molecules are exclusively found at the surface. It has been argued that the slightly larger cavity required by the Sr-HCN molecule inside a helium droplet makes it more heliophobic than the bare atom. Interestingly, it has been shown that the Sr-HCN complex is trapped in an energy minimum at the surface and that upon vibrational excitation the barrier that prevents the system from solvation can be overcome. Similar, for the Ca-HCN system vibrational excitation can transfer population between the two locations.  \\ 
Beyond calculations for Mg dimers in helium droplets\cite{hernando2008density} a few other theoretical studies on this topic have been reported. Among them is the predicted formation of a quantum gel structure for Ne doped superfluid (bulk) helium,\cite{eloranta2008self} which is expected to consist of Ne atoms surrounded by He solvent shells in a crystalline structure. Other calculations suggest that He droplets with a Xe atom at the center and a Rb atom at the surface support a situation with separated atoms.\cite{poms2012helium}  \\
Here, we report the existence of helium droplets that accommodate spatially separated Rb and Sr atoms. Evidence for such a situation is based on the recording of the Sr-He$_{\text{N}}$ $5s5p$ $^1$P$^o_1\leftarrow 5s^2$ $^1$S$_0$ transition, which can be monitored at the RbSr, the Rb as well as the Sr$_2$ mass window (employing resonant two-photon ionization spectroscopy), unambiguously showing that the Sr atom was separated from the other atom at the moment of excitation. Before we discuss the corresponding excitation spectra we report on new insights into the Sr-He$_{\text{N}}$ $5s5p$ $^1$P$^o_1$ system, for which we show that an excitation does not necessarily lead to a desorption. Surprisingly, our results also suggest that a fraction of the Sr dopant atoms reside inside the helium droplets.   \\
The experimental methods and setup have been described in more detail elsewhere \cite{Nagl2007,Nagl2008,Lackner2011}. In brief, helium droplets are produced by expanding helium gas under high pressure (60\,bar) through a cold (15\,K) 5\,$\mu$m nozzle into vacuum. At these conditions, helium droplets with a maximum of the log-normal droplet size distribution at $\hat{N} = 6000$ are generated.\cite{Toennies2004,Callegari2011} The helium droplet beam is first guided through a pickup cell loaded with Rb and, subsequently, through a second cell loaded with Sr.\cite{lackner2014helium} \\
A pulsed Radiant Dyes RD-EXC 200 XeCl laser (26 ns pulse duration, 100 Hz) is employed to pump a Lambda Physik FL 3002 dye laser (bandwidth 0.2\,cm$^{-1}$) operated with Coumarin 2, which is used to excite the Sr $5s5p$ $^1$P$^o_1$ state (at 21699\,cm$^{-1}$ in the bare Sr atom).\cite{Nist2016} The peak intensity of the laser dye was about 0.5\,mJ, focused to a spot size of $\sim$1\,mm$^2$. A resonant two-photon ionization (R2PI) scheme is employed, where a second photon is provided by the XeCl pump laser ($\sim$0.3\,mJ), which is spatially overlapped with the dye laser. The 308 nm (32468\,cm$^{-1}$) photon is required considering that the ionization threshold of Sr corresponds to 45932\,cm$^{-1}$.\cite{Nist2016} Ions are detected mass selectively as a function of the dye laser wavelength using a time of flight (TOF) mass spectrometer. The presented spectra are monitored at mass windows corresponding to the most abundant isotopes $^{88}$Sr (82.6\%), $^{85}$Rb (72.2\%) and $^{85}$Rb$^{88}$Sr.\cite{CRC84}  \\
Dispersed fluorescence spectra have been recorded using a grating monochromator with an attached CCD camera. The $457.9$ nm line (21839\,cm$^{-1}$) of an Ar-ion laser has been used to populate the Sr-He$_{\text{N}}$ $5s5p$ $^1$P$^o_1$ state.  \\ 
\begin{figure}
\centering
\includegraphics[width=8 cm]{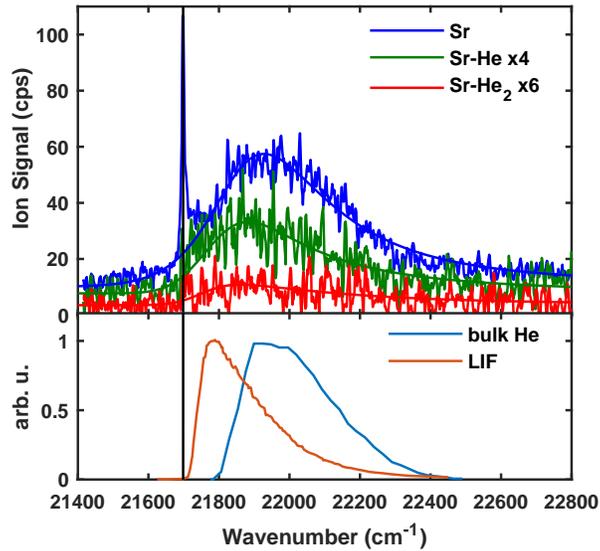}
\caption{Upper panel: Excitation spectrum of the $5s5p$ $^1$P$^o_1\leftarrow 5s^2$ $^1$S$_0$ transition of Sr-He$_{\text{N}}$ recorded at mass windows corresponding to $^{88}$Sr (blue), $^{88}$Sr-He (green) and $^{88}$Sr-He$_2$ (red). The bare Sr atom $5s5p$ $^1$P$^o_1\leftarrow 5s^2$ $^1$S$_0$ transition is only visible in the $^{88}$Sr spectrum and can be seen at 21699\,cm$^{-1}$.\cite{Nist2016} Lower panel: Literature Sr-He$_{\text{N}}$ LIF (orange) and bulk superfluid He spectra (black) adopted from ref \citenum{stienkemeier1997alkaline} and ref \citenum{bauer1990laser} , respectively.
} 
\label{fig1}
\end{figure}
The $5s5p$ $^1$P$^o_1\leftarrow 5s^2$ $^1$S$_0$ transition is the lowest optically allowed Sr transition and has been investigated previously for Sr-He$_{\text{N}}$ using laser induced fluorescence (LIF) spectroscopy.\cite{stienkemeier1997alkaline,hernando2007structure} Previous results favored a scenario where Sr atoms reside in a dimple at the surface of helium nanodroplets.\cite{ren2007surface,stienkemeier1997alkaline,hernando2007structure} The LIF spectra, which is shown in the lower panel of Figure~\ref{fig1},  (adopted from ref \citenum{stienkemeier1997alkaline}) exhibits a blue-shifted (80\,cm$^{-1}$)\cite{hernando2007structure} and broadened structure. Note that the shift and the width of the LIF spectrum is independent of the droplet size for mean He$_{\text{N}}$ sizes larger than $N \sim 3000$.\cite{stienkemeier1997alkaline} R2PI spectroscopy provides additional insights by adding the component of mass sensitivity. The corresponding spectra recorded at the mass windows of $^{88}$Sr (blue), $^{88}$Sr-He (green) and $^{88}$Sr-He$_2$ (red) are shown in Figure~\ref{fig1}, top panel. The Sr spectrum, shown in blue, features the narrow bare atom $5s5p$ $^1$P$^o_1\leftarrow 5s^2$ $^1$S$_0$ transition as well as a broad band blue-shifted by $230\pm25$\,cm$^{-1}$. The band is also visible in the Sr-He spectrum but blue-shifted by about  $185\pm25$\,cm$^{-1}$. The difference between Sr and Sr-He ion yield spectra indicates that the probability for the formation of an exciplex (an excited Sr atom with attached He atoms) decreases with increasing photon energy. Note that a faint band is also visible in the Sr-He$_2$ spectrum, shown in red. \\
Compared to the LIF spectroscopy results\cite{stienkemeier1997alkaline,hernando2007structure} the blue shift and broadening exhibited by the band in the R2PI spectra are much larger. The recorded R2PI more closely resembles the spectrum of Sr atoms in bulk superfluid He\cite{bauer1990laser}, shown in the bottom panel of Figure~\ref{fig1}, except for the gap between the rising edge and the bare atom line. Thus, in order to explain the recorded spectrum one has to consider both cases: Solvated Sr atoms giving rise to the strongly broadened and blue shifted part of the spectrum but also a contribution of surface site Sr atoms that explain the weakly shifted signal in the gap between the bulk spectrum and the bare atom line. Previously, such a situation has been described for Ca-HCN doped He droplets.\cite{douberly2010hcn} In fact, the integrated R2PI spectrum (sum over Sr, Sr-He and Sr-He$_2$ spectra) can be described by a superposition of the the spectrum of Sr solvated in bulk superfluid He and the literature Sr-He$_{\text{N}}$ LIF spectrum with a ratio of 7/5, respectively, providing an estimation for the ratio between solvated and surface site Sr atoms. The simulated spectrum is shown in the Supporting Information.  \\
This finding is supported by calculations of the Sr-He$_{\text{N}}$ solvation energy as a function of the distance of the dopant from the droplet center using the Sr-He pair potentials from Lovallo \textit{et al.}\cite{lovallo2004accurate}. The details of the calculation can be found in the Supporting Information. The unexpected but intriguing result suggests two minima in the solvation energy, one at the surface and a second in the center of the droplet both separated by a small barrier. Thus, also from this calculation it is concluded that Sr atoms are not only situated at the surface of the droplet, they can also reside inside the droplet. \\
\begin{figure}
\centering
\includegraphics[width=8 cm]{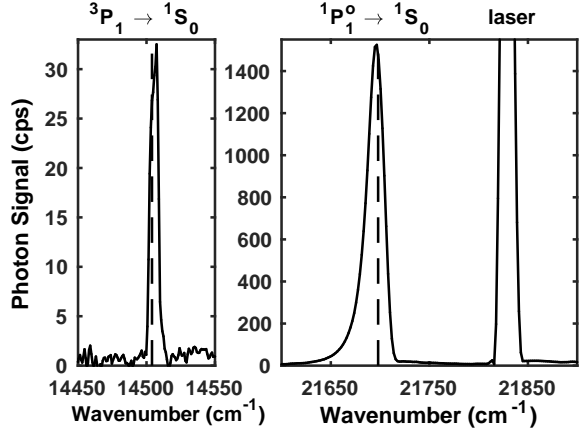}
\caption{Dispersed fluorescence spectrum of Sr-He$_{\text{N}}$ for the energy windows containing the $5s5p$ $^1$P$^o_1\rightarrow 5s^2$ $^1$S$_0$ and 5s5p $^3$P$^o_1\rightarrow 5s^2$ $^1$S$_0$ lines. Stray light caused by the excitation laser can be seen at 21839\,cm$^{-1}$. The $5s5p$ $^1$P$^o_1\rightarrow 5s^2$ $^1$S$_0$ emission exhibits a characteristic asymmetric broadening, suggesting that a substantial amount of Sr atoms undergoes bound-bound transitions and stays attached to the droplet upon excitation.     
}
\label{fig2}
\end{figure}
Figure~\ref{fig2} shows dispersed fluorescence spectra of Sr-He$_{\text{N}}$ upon excitation at 21839\,cm$^{-1}$, close to the maximum of the peak in the Sr-He ion yield spectrum in Figure~\ref{fig1}. The bare Sr atom $5s5p$ $^1$P$^o_1\rightarrow 5s^2$ $^1$S$_0$ emission line at 21699\,cm$^{-1}$ is indicated by a vertical dashed line.\cite{Nist2016} It can be seen that the emission line exhibits an asymmetric broadening towards lower photon energies, the spectroscopic fingerprint of a bound-bound transition. Very similar spectra have been reported for Rb-He$_{\text{N}}$\cite{aubock2008electron} and Cs-He$_{\text{N}}$\cite{theisen2011cs} at the D1 line, which served as evidence that these atoms stay bound to the droplet surface upon excitation of the respective transition. Consequently, we conclude that the Sr atom stays bound to the droplet with a certain probability upon excitation at 21839\,cm$^{-1}$. This also suggests that the LIF spectrum from ref. \citenum{stienkemeier1997alkaline} is biased towards lower photon energy as atoms that remain bound to the surface can contribute multiple times to the signal (the life-time of the $5s5p$ $^1$P$^o_1$ Sr state is 4.98\,ns)\cite{Nist2016}.  \\     
This conclusion is further supported by the detection of large Sr-He$_{\text{N}}$ complexes (with N $>$ 1000) when ionizing via the Sr-He$_{\text{N}}$ $5s5p$ $^1$P$^o_1\leftarrow 5s^2$ $^1$S$_0$ transition at 21830\,cm$^{-1}$(shown in the Supporting Information). Note that this has also been observed for Ba-He$_{\text{N}}$, which also remains bound to the droplet upon excitation of the lowest $^1$P$^o_1\leftarrow$ $^1$S$_0$ transition.\cite{loginov2012spectroscopy,zhang2012communication}. \\
A further interesting finding, which also indicates that the Sr atom tends to stay attached to the droplet, is shown in the left panel of Figure~\ref{fig2}. A fraction of the Sr atoms undergoes relaxation into the metastable 5s5p $^3$P$^o_J$ state manifold. Fluorescence light can be observed from the 5s5p $^3$P$^o_1\rightarrow 5s^2$ $^1$S$_0$ intercombination line. Note that the respective transition probability to the ground state is about 4 orders of magnitude lower than for the $5s5p$ $^1$P$^o_1$ state.\cite{Nist2016} Our observation demonstrates that droplet mediated (non-radiative) relaxation is an important process for Sr-He$_{\text{N}}$, in particular, for the fully solvated fraction of atoms\cite{kautsch2013electronic,Messner2018,loginov2007excited}.  \\
This can explain the difference between LIF and R2PI spectra: First, Sr atoms trapped in the metastable 5s5p $^3$P$^o$ states (but also in the 5s4d $^3$D and $^1$D states that are potentially populated) can be ionized with a single 308 nm photon and thus contribute to the R2PI signal. However, the contribution of these relaxed atoms to the LIF signal is negligible. As we expect the droplet mediated relaxation process to be more efficient for solvated atoms, we conclude that the LIF signal is dominated by surface site or desorbed atoms that did not undergo a relaxation. In contrast, the R2PI spectrum includes signal from both solvated and surface site Sr atoms.  \\
\begin{figure}
\centering
\includegraphics[width=8 cm]{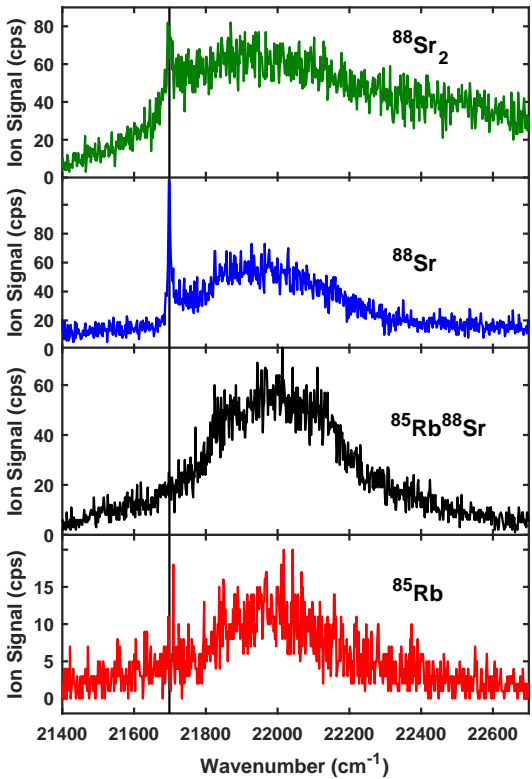}
\caption{Excitation spectra of the Sr-He$_{\text{N}}$ $5s5p$ $^1$P$^o_1\leftarrow 5s^2$ $^1$S$_0$ transition recorded at mass windows corresponding to $^{85}$Rb (red), $^{85}$Rb$^{88}$Sr (black), $^{88}$Sr (blue) and $^{88}$Sr$_2$ (green). The bare Sr atom $5s5p$ $^1$P$^o_1\leftarrow 5s^2$ $^1$S$_0$ transition is marked by a vertical black line at 21699\,cm$^{-1}$\cite{Nist2016}. 
}
\label{fig3}
\end{figure}
Excitation spectra recorded for the mass channels corresponding to $^{85}$Rb (red), $^{85}$Rb$^{88}$Sr (black), $^{88}$Sr (blue) and $^{88}$Sr$_2$ (green) are shown in Figure~\ref{fig3}. The blue $^{88}$Sr spectrum serves as a reference for the $5s5p$ $^1$P$^o_1\leftarrow 5s^2$ $^1$S$_0$ transition of the Sr-He$_{\text{N}}$ system. Strikingly, when the droplets are co-doped with Rb atoms, it becomes evident that both the Rb and RbSr spectra exhibit the droplet broadened $5s5p$ $^1$P$^o_1\leftarrow 5s^2$ $^1$S$_0$ transition. %Note that a contamination of the shown spectra by other isotopes or isotope combinations can be excluded as the shown mass windows only contain the respective species.\cite{krois2014characterization} 
The effect is best seen in the black RbSr spectrum where the Sr-He$_{\text{N}}$ $5s5p$ $^1$P$^o_1\leftarrow 5s^2$ $^1$S$_0$ transition can be unambiguously identified. Even though RbSr transitions are present in the respective energy regime,\cite{lackner2014helium,krois2014characterization} we exclude the possibility that one of them has the exact same shape and position as the Sr-He$_{\text{N}}$ $5s5p$ $^1$P$^o_1\leftarrow 5s^2$ $^1$S$_0$ transition. Moreover, evidence for the feature in Figure~\ref{fig3} can also be found in the spectrum shown in refs. \citenum{lackner2014helium} and \citenum{krois2014characterization}, however, in these works the corresponding feature eluded from an assignment. \\
The only possible scenario that explains the observed signals is that the atoms are initially separated and do not form a molecular bond upon pickup. However, we emphasize that this is not the only possible pathway, and on a significant fraction of helium droplets the atoms do form a molecular bond as demonstrated by the recording of molecular RbSr spectra.\cite{lackner2014helium,krois2014characterization} Furthermore, we note that a similar observation has also been reported for RbCa spectra\cite{pototschnig2015investigation}: In this case a weak signal that followed the droplet broadened Rb-He$_{\text{N}}$ $4d~ ^2\text{D} \leftarrow 5s~ ^2\text{S}$ transition at the RbCa mass window has been observed.  \\
The results indicate the following scenario: A large fraction of Sr atoms does not immediately desorb upon excitation, moreover, photo-excitation triggers a reaction during which the molecular bond is formed. Presumably, this scenario is more favored by a combination of surface site Rb and fully solvated Sr atoms. The energy released by this process leads to a desorption of the molecule, which is subsequently ionized and detected in the mass spectrometer. Interestingly, the Rb spectrum, shown in red, exhibits the same feature, indicating that this process can also lead to the desorption of bare Rb atoms. \\
If this pickup pathway is indeed a more general phenomenon, an obvious test is the spectroscopy of Sr dimers, shown in the top panel of Figure~\ref{fig3}. Note that for this experiment only the dye laser has been used and the pickup oven temperature has been raised in order to increase the probability for the pickup of two Sr atoms by one droplet. Indeed, it becomes evident that the spectrum exhibits also a feature that can be attributed to the Sr-He$_{\text{N}}$ $5s5p$ $^1$P$^o_1\leftarrow 5s^2$ $^1$S$_0$ transition. Surprisingly, the spectrum also exhibits a peak reminiscent of the bare atom $5s5p$ $^1$P$^o_1\leftarrow 5s^2$ $^1$S$_0$ transition.  \\  
The previously best documented case of a dopant for which separated atoms in helium droplets were suggested is the Mg atom.\cite{przystawik2008light} A modulation of the long-range part of the weak bond of the Mg dimer ($D_e = 430$\,cm$^{-1}$)\cite{li2010van} has been argued to be responsible for the formation of a metastable, foam-like structure. The same argument may also be claimed for the weakly bound high-spin species Na$_2$ and K (e.g. $D_e = 175$\,cm$^{-1}$ for the Na$_2$ triplet ground state).\cite{friedman1992reexamination} For these dopants, a K atom fluorescence line allowed to record a Na dimer transition, which suggested the presence of separated species on the droplet surface and the collisional transfer of energy between the excited dimer and the ground state K atom.\cite{Higgins1998} We note that compared to these species, the binding energies of the RbSr X$^2\Sigma^+$ (1273\,cm$^{-1}$)\cite{pototschnig2014ab} and Sr$_2$ X$^1\Sigma^+_g$ (1070\,cm$^{-1}$)\cite{czuchaj2003valence} states are much stronger, suggesting that the described alternative pickup pathway is not restricted to very weakly bound molecules.  \\
In conclusion, we demonstrated that there is a second important but commonly neglected pathway that can be undertaken by dopants upon the pickup by a helium droplet. On the example of Rb and Sr doped droplets it is shown that there is a considerable amount of atoms that stay separated upon pickup in addition to the formation of RbSr molecules. This suggests that the presence of spatially separated atoms is not an exotic phenomenon restricted to very weakly bound complexes\cite{przystawik2008light,poms2012helium} but much rather becomes also relevant for more strongly bound species if they occupy sites in/on the droplet that are separated by a barrier.\\
Our results suggest another surprising phenomenon that has, so far, never been considered for atomic dopants: In addition to the commonly assumed situation of Sr atoms on the surface of the helium nanodroplets, our results suggest the presence of fully immersed Sr atoms in a fraction of the droplets. This explains the observed differences between LIF and R2PI spectra and is supported by calculations of the solvation energy of the Sr atom as a function of its distance from the droplet center. In this regard, it would be very interesting to explore other alkaline-earth metal dopants, for which a similar scenario may be expected.  \\    
%The Sr atom has been a good choice to demonstrate this process because, as shown by the recorded dispersed fluorescence and R2PI spectra, the atom does not immediately desorb upon photo-excitation and can react with the second educt to form a molecule. \\
%Mg doped droplets were always considered as an unusual system, Mg atoms were claimed to form foam-like structures inside helium dropelts.\cite{przystawik2008light} 
It is only recently that indications for the presence of spatially separated atoms on or inside a droplet are becoming more frequent, as the results for Al-He$_{\text{N}}$\cite{krasnokutski2011low,krasnokutski2015resonant,spence2014formation} or the examples of atom-like transitions recorded for Au-He$_{\text{N}}$\cite{Messner2018} and Cr-He$_{\text{N}}$\cite{kautsch2015photoinduced} oligomer mass channels suggest. In the synthesis of metal cluster and nanoparticles coagulation at multiple aggregation centers is a well known phenomenon\cite{volk2016impact,haberfehlner2015formation}. The fact that even on the smallest scale we now observed the formation of separated species may put fundamental aspects of the cluster aggregation process into a new light.  \\

\acknowledgement

The authors gratefully acknowledge support by the Austrian Science Fund (FWF) under grant P30940-N36 as well as support from NAWI Graz. We thank G\"unter Krois for his help and support during the experiments and Ralf Meyer for help and support with the calculations. F.L. acknowledges support by the Austrian Science Fund (FWF) via an Erwin Schr\"odinger Fellowship under Grant No. J 3580-N20. 

\section{Supporting Information}

Mass spectra of large Sr-He$_{\text{N}}$ complexes and calculation of solvation energies and a simulation of the R2PI spectrum using literature LIF and bulk superfluid He spectra.

%\bibliographystyle{achemso} 
%\bibliography{achemso-demo}
\bibliography{RbSr}

\end{document}